\documentclass[journal,transmag]{IEEEtran}
\usepackage{cite}

\ifCLASSINFOpdf
   \usepackage[pdftex]{graphicx}
   \graphicspath{{figpdf/}}
   \DeclareGraphicsExtensions{.pdf}
\else
\fi
\usepackage{amsmath}
\usepackage{amsfonts}
\usepackage{siunitx}
\newcommand{\uH}{\si{\micro\henry}}
\newcommand{\mT}{\si{\milli\tesla}}
\newcommand{\kHz}{\si{\kilo\hertz}}
\newcommand{\mgpmL}{\si[per-mode=symbol]{\milli\gram\per\milli\liter}}
\newcommand{\ug}{\si{\micro\gram}}
\newcommand{\ms}{\si{\milli\second}}
\usepackage{array}
\usepackage{multirow}
\usepackage{rotating}

\ifCLASSOPTIONcompsoc
 \usepackage[caption=false,font=normalsize,labelfont=sf,textfont=sf]{subfig}
\else
 \usepackage[caption=false,font=footnotesize]{subfig}
\fi
\usepackage{url}

\begin{document}
%
\title{Vector Modulator Based Active Compensation\\ of Direct Feedthrough in Magnetic Particle Imaging}


\author{\IEEEauthorblockN{Bilal Tasdelen\IEEEauthorrefmark{1,2},
Mustafa Utkur\IEEEauthorrefmark{1,2},
Asli Alpman\IEEEauthorrefmark{1,2}
Can Baris Top\IEEEauthorrefmark{3}, 
Ergin Atalar\IEEEauthorrefmark{1,2,4},~\IEEEmembership{Senior Member,~IEEE},\\ and
Emine Ulku Saritas\IEEEauthorrefmark{1,2,4},~\IEEEmembership{Member,~IEEE}}
\IEEEauthorblockA{\IEEEauthorrefmark{1}Department of Electrical and Electronics Engineering, Bilkent University, Ankara, Turkey}
\IEEEauthorblockA{\IEEEauthorrefmark{2}National Magnetic Resonance Research Center (UMRAM), Bilkent University, Ankara, Turkey}
\IEEEauthorblockA{\IEEEauthorrefmark{3}ASELSAN Research Center, Ankara, Turkey}
\IEEEauthorblockA{\IEEEauthorrefmark{4}Neuroscience Program, Aysel Sabuncu Brain Research Center, Bilkent University, Ankara, Turkey}
\thanks{Manuscript received XX May 2024; revised XX September 2024; accepted XX October 2024. This work was supported by the Scientific and Technological Research Council of Turkey under Grant TUBITAK 120E208. The authors would like to thank Ecrin Yagiz for her help during the MPS experiments. Preliminary versions of this work were presented at the 10\textsuperscript{th} International Workshop on Magnetic Particle Imaging (IWMPI 2020) and the Annual Meeting of International Society for Magnetic Resonance in Medicine (ISMRM 2021). \textit{(Corresponding authors: Bilal Tasdelen, Emine U. Saritas)}

Bilal Tasdelen was with Department of Electrical and Electronics Engineering, Bilkent University, Ankara, Turkey. He is now with Department of Electrical and Computer Engineering, University of Southern California, Los Angeles, CA, USA. (e-mail: billtasdelen@gmail.com)

Mustafa Utkur was with Department of Electrical and Electronics Engineering, Bilkent University, Ankara, Turkey. He is now with Department of Radiology, Boston Children’s Hospital and Harvard Medical School, Boston, MA, USA.

Asli Alpman was with Department of Electrical and Electronics Engineering, Bilkent University, Ankara, Turkey. She is now with Department of Electrical Engineering and Computer Sciences, University of California, Berkeley, Berkeley, CA, USA. 

Can Baris Top is with ASELSAN Research Center, Ankara, Turkey. (e-mail: cbtop@aselsan.com.tr)

Ergin Atalar is with Department of Electrical and Electronics Engineering, Bilkent University, Ankara, Turkey. (e-mail: ergin.atalar@bilkent.edu.tr)

Emine U. Saritas Department of Electrical and Electronics Engineering, Bilkent University, Ankara, Turkey. (e-mail: saritas@ee.bilkent.edu.tr)}}

%



\IEEEtitleabstractindextext{%
\begin{abstract}
In magnetic particle imaging (MPI), simultaneous excitation and signal acquisition leads to direct feedthrough interference. While this interference can be mitigated up to some extent with passive compensation, its time-varying nature necessitates active compensation methods to achieve the sensitivity levels needed for applications such as stem cell tracking. We have recently proposed an active compensation technique for MRI, which uses a vector modulator and a lookup-table-based algorithm for reducing the direct feedthrough in the analog domain. Here, we adapt this technique to MPI, demonstrating a successful recovery of the fundamental frequency and a significant increase in detection sensitivity.
\end{abstract}

}

\maketitle


%
\IEEEpeerreviewmaketitle

\section{Introduction}

\IEEEPARstart{M}{agnetic} Particle Imaging (MPI) is a rapidly developing modality that images the spatial distribution of magnetic nanoparticles (MNPs) by exploiting their nonlinear magnetization curves \cite{gleich_tomographic_2005, saritas_magnetic_2013}. MPI has excellent potential for numerous clinical applications such as angiography \cite{rahmer_interactive_2017}, stem cell imaging\cite{zheng_quantitative_2016}, viscosity mapping\cite{stehning_simultaneous_2016, utkur_relaxation-based_2017}, functional imaging\cite{mason_design_2017}, and inflammation imaging \cite{weaver_identifying_2020, chandrasekharan_non-radioactive_2021}. 

In MPI, a sinusoidal drive field (DF) is applied to excite the MNPs. 
Since MNPs align with the applied DF almost instantaneously, excitation and signal acquisition are inherently simultaneous in MPI. Furthermore, 
the pulse transmitted from the drive coil couples to the receive coil and induces a direct feedthrough signal \cite{lu_linearity_2013}. Because this direct feedthrough can be orders of magnitude larger than the MNP signal, it can easily saturate the preamplifier and reduce the dynamic range, jeopardizing MPI’s promise of high sensitivity imaging \cite{goodwill_x-space_2010}. As a first measure of reducing the direct feedthrough, gradiometer receive coils are widely utilized in MPI to minimize the mutual inductance between the drive and the receive coils \cite{karp_unidirectional_1980,  utkur_comparison_2015, graeser_towards_2017, cagil_design_2020}. While gradiometer receive coils passively reduce the direct feedthrough signal, their tuning is prone to degradation and drift due to factors such as vibration and heating. As a result, even after passive cancellation, the residual feedthrough can easily exceed the MNP signal. A typical secondary measure is to acquire background measurements to be subtracted from the imaging signal. Although this simple approach can further reduce the direct feedthrough, system drift during long scan durations can cause significant temporal variations in its efficacy \cite{pantke_multifrequency_2019, kurt_partial_2020, knopp_efficient_2021}. Due to these difficulties, the fundamental frequency is almost always filtered out of the received signal \cite{lu_linearity_2013}.

Another approach for reducing the direct feedthrough is via active compensation (AC), where the time-dependent direct feedthrough is actively monitored and subtracted from the received signal at the analog stage. Time-dependent direct feedthrough (also known as self-interference in telecommunications) is a challenging but ubiquitous issue. Therefore, AC is an active and vital area of research not just in MPI \cite{zheng_high-power_2013, pantke_multifrequency_2019}, but in other fields such as telecommunications (as full-duplex radio problem)\cite{bharadia_full_2013, saheed_tijani_low-power_2017, van_den_broek_-band_2015, ramakrishnan_design_2017,zhou_low-noise_2014} and MRI (as concurrent excitation and acquisition problem)  \cite{idiyatullin_continuous_2012, tasdelen_dynamic_2019, tasdelen_analysis_2021, ozen_active_2015, ozen_vivo_2018, sohn_vivo_2016, salim_active_2018}. 
Indeed, the ability to actively reduce the direct feedthrough would be extremely useful for almost all applications of MPI. In imaging, it can enable the fundamental frequency to be preserved, which in turn can improve the image signal-to-noise ratio (SNR) and thereby the detection sensitivity \cite{pantke_multifrequency_2019}. In addition, preserving the fundamental frequency can circumvent the image artifacts associated with the recovery of the fundamental frequency in x-space reconstruction  \cite{kurt_partial_2020}. Furthermore, the increase in SNR can enable imaging at lower DF strengths. As a result, cheaper systems can be built with less stringent power and cooling requirements and with less emphasis on building a carefully tuned gradiometer coil, increasing the availability and robustness of MPI.

Previously in MPI, AC was performed by utilizing a second independent signal source \cite{zheng_high-power_2013,pantke_multifrequency_2019}. Although this approach provides simplicity and flexibility for creating the cancellation signal, it also requires a precise knowledge of the DF waveform. Furthermore, in the case that transmit noise is dominant, addition of an independent transmit source can significantly increase the received noise level \cite{ozen_active_2015}. Furthermore, if the DF gets distorted 
due to heating, vibrations or other non-linear distortions, this method will fail to compensate. 

In this work, we propose a vector modulator (VM) based AC method to address the direct feedthrough problem in MPI. The proposed VM is integrated to the MPI system and is digitally controlled, empowering the MPI system with a fully automated AC process that does not require any manual interventions. In addition, we propose a fast look-up table-based algorithm that automatically and dynamically compensates for the time-varying direct feedthrough. The proposed AC method and the algorithm are demonstrated via extensive simulations and experiments in a magnetic particle spectrometer (MPS) setup. The simulation results show that the additional feedthrough cancellation reaches up to 70~dB at high direct feedthrough levels. In addition, the experiments show that the proposed method provides 39~dB of additional cancellation in 500 milliseconds, successfully recovering the MNP signal at the fundamental frequency and significantly improving the detection sensitivity. 

\section{Methods}
\subsection{Direct Feedthrough Signal}

The direct feedthrough signal in MPI stems from the non-zero mutual inductance between the closely wound DF coil and receive coil. Usually, the receive coil is configured as a gradiometer coil to reduce the mutual inductance. Theoretically, for the case of a perfect construction, it is possible to zero out the mutual inductance. In practice, however, it is extremely challenging to achieve this goal due to possible constructional flaws. 

Regardless of the configuration, the effect of the mutual inductance is considered as the coupling of the DF waveform to the receive side, attenuated and phase shifted. If we consider a sinusoidal DF with a center frequency $f_0$ and envelope $A(t)$, the received direct feedthrough at $f_0$ can be written as:
\begin{equation}\label{eq:ft_formula}
	s(t) = k(t) \cdot A(t) \cdot \sin(2\pi f_0 t + \phi(t)),
\end{equation}
where $k(t)$ and $\phi(t)$ are the attenuation coefficient and the phase shift stemming from mutual coupling, respectively, which can both vary with time. The main reasons for this time-variance of the mutual inductance are the heating of the DF coil due to applied power, temperature fluctuations in the environment, vibrations, or other small movements in the system. It should be noted that the system becomes more susceptible to these effects as the mutual inductance becomes smaller.

One approach to reduce this direct feedthrough is actively removing it from the received signal using the knowledge of the DF signal. This can be done either in the digital or in the analog domain. Digital cancellation of the feedthrough is advantageous in terms of both system complexity and cost, as it is not encumbered by the challenges of hardware design. However, it fails to accommodate for several outstanding issues such as the reduction of the receive chain dynamic range, unforeseen deviations of the DF due to hardware imperfections caused by amplifier nonlinearity, and interference from external sources into the DF signal.

\subsection{Active Compensation Circuitry}
\label{subsection_AC_circuit}

\begin{figure}[!t]
	\centering
	\includegraphics[width=0.9\columnwidth]{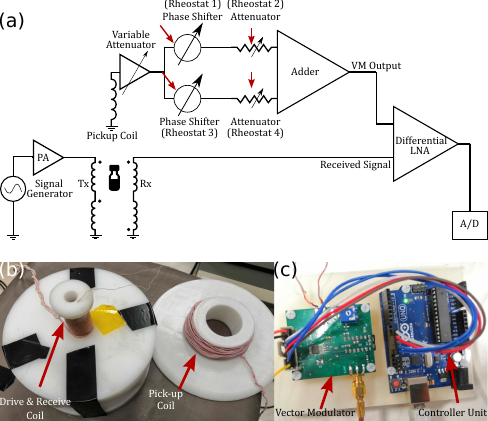}
	\caption{(a) System diagram of the setup and the vector modulator. Red arrows show the places where digital rheostats are deployed. 
	(b) Photograph of the gradiometer coil and the pick-up coil.
	(c) Photograph of the vector modulator and the controller unit.}
	\label{fig_diag}
\end{figure}

In this work, an AC scheme is proposed that estimates the parameters $k$ and $\phi$ and creates a replica of the direct feedthrough signal to remove it in the analog domain. Here, both $k$ and $\phi$ are assumed to be stable during short intervals of signal acquisition, but may drift over time. The aforementioned replica of the direct feedthrough will be referred to as the ``compensation signal". To adjust the compensation signal according to the parameters $k$ and $\phi$, we propose a VM circuit that can modify both the amplitude and phase of its input, as shown in Fig.~\ref{fig_diag}. This circuit contains two branches, where each branch employs a phase shifter and an attenuator: The phase shifter provides up to $\pi$ relative phase between the input and the output, whereas the attenuator adjusts the relative amplitude of the compensation. As a result, the compensation provided by each branch approximately covers a complex half-plane in phasor domain. These two branches are then combined with weighting, enabling approximately full coverage of the entire complex plane. Finally, the resulting compensation signal is subtracted from the received signal.

The phase shifters in Fig.~\ref{fig_diag} were realized as first-order all-pass filters \cite{maheshwari_analog_2007}. The relative phase of an all-pass filter in a low-pass configuration at the inverting input can be calculated as:

\begin{equation} \label{eq:allpasslp}
\angle H(i \omega) = -2 \: \arctan (\omega R C),
\end{equation}
where $\omega$ is the radial frequency. In a high-pass configuration, on the other hand, there is an additional $\pi$ phase in Eq.~\ref{eq:allpasslp}. Hence, by fixing C and sweeping R via a rheostat, we can cover phases from 0 to $-\pi$ with a low-pass configuration and from 0 to $\pi$ with a high-pass configuration. Here, for a given set of available R values of a rheostat, C needs to be chosen to ensure that a sufficient range of phases are covered while being sampled densely enough. Moreover, since the phase is frequency dependent, the operating regime needs to be taken into account. 

In this work, linearly varying 8-bit digital rheostats (MCP4352, Microchip) with 5~k$\Omega$ maximum resistances were used for implementing the phase shifters and the attenuators, and C was chosen as 22~nF to provide the necessary phase coverage at $f_0$ = 10~kHz.
The attenuators controlled the gain/attenuation for each branch of the weighted adder, which was implemented as an inverting Op-Amp (TL074) summing amplifier. An external PC calculated the digital codes for the rheostats, and an Arduino microcontroller translated these codes into serial commands to configure the rheostat states. 

As shown in Fig.~\ref{fig_diag}a, at the input of the VM circuit, the feedthrough signal was sensed from the DF coil via a pick-up coil with an inductance of 313~\uH. This relatively large inductance provided high sensitivity to the DF signal while ensuring minimal interaction with the receive coil. An inverting Op-Amp (TL074) amplifier was used to scale the gain of the VM and isolate the input impedance due to the pick-up coil from the phase shifters. Finally, the output of the VM was subtracted from the received signal using the differential input of a low noise amplifier (LNA; Stanford Research Systems SR560). 

\subsection{Active Compensation Algorithm}
\label{subsection_AC_algo}

To boost the performance of the VM, we propose an AC algorithm that can control the VM via its digital inputs (i.e., the wiper states of the 8-bit digital rheostats) and minimize the residual feedthrough. Such an algorithm needs to be fast enough to accommodate a time-varying feedthrough (i.e., system drift), while being precise enough to minimize the residual feedthrough to a sufficiently low level.
A non-linear integer optimization algorithm can be run to determine the digital inputs that minimize the residual feedthrough, by measuring only the signal at the output of the differential LNA in Fig.~\ref{fig_diag}. Importantly, this measurement does not require the explicit knowledge of either the feedthrough or the VM output. 
However, such optimization algorithms typically require many consecutive measurements to converge, leading to a very time-consuming operation, which is especially problematic if the feedthrough drifts quickly over time. To avoid this convergence issue, we adopt a fast, system characterization based approach in this work.

\textit{1) Look-up Table for System Characterization}: In theory, the output phase and amplitude of the VM for a given input signal can be estimated, since the transfer function of this circuit can be derived using the component values for fixed digital inputs. However, a better characterization of the system is necessary for various reasons such as the tolerances in the circuit components, which can limit the performance of the VM. Here, the most prominent nonidealities are due to the values of the resistors and wiper resistances in the digital rheostats, which may have up to 1\% error as stated in the datasheet. Another prominent reason necessitating a better characterization is the system drift. To mitigate these issues, we propose a measurement based characterization procedure. 

\begin{figure}[!t]
	\centering
	\includegraphics[width=\columnwidth]{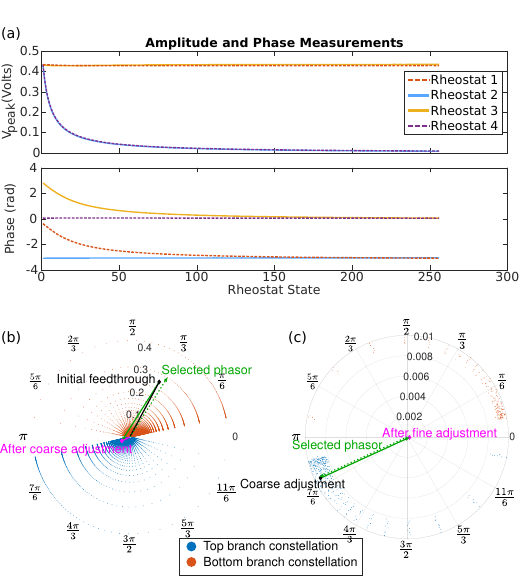}
	\caption{(a) Amplitude and phase measurements, in the case where the state of each rheostat is swept one by one, while the other rheostats are disconnected. With each rheostat having $N$ = 256 states, a total of $4N = 2^{10}$ measurements were conducted by digitally controlling the rheostat states. These measurements were then used to create the look-up table constellations that map the phasor of the output contribution by each branch.
	Graphical representation of the (b) coarse adjustment and (c) fine adjustment steps of the LUT algorithm in phasor domain. For each step, the black arrows represent the measured initial feedthrough phasor, the green dashed arrow shows the selected phasor for subtraction, the solid green arrow shows the subtraction operation. The purple arrow shows the residual feedthrough after each adjustment step. Note the scale difference along the radial direction between the coarse and fine adjustment steps in (b) and (c). These plots show highly undersampled versions of the estimated constellation, which contained $ 2^{17}$ points for each step (i.e., $N^2 = 2^{16}$ points each for top and bottom branch constellations). }
	\label{fig_algo}
\end{figure}

Here, we refer to the entire possible input-output mapping of the VM in phasor domain as the ``constellation". Assuming that each rheostat has $N \in \mathbb{N}$ different wiper states (i.e. $N$ possible resistance values), $N^4$ measurements would be needed to fully characterize the constellation given that the circuit contains 4 rheostats. However, if we assume that the two branches of the VM are decoupled from each other (a reasonable assumption considering the Op-Amps at the input/output of each branch), the overall transfer function acting on the input is equal to the vector summation of the transfer functions of each branch. Hence, to characterize the modulator, one can disconnect one of the branches and measure the amplitude and phase shift with respect to the input while sweeping the rheostats in the connected branch, covering the constellation in $2N^2$ measurements instead. 

In this work, we propose going one step further and sweeping each rheostat one by one. First, the bottom branch is disconnected, and the attenuator rheostat on the top branch is kept at a fixed wiper state while the phase shifter rheostat wiper is swept. Next, for the same branch, the phase shifter rheostat is kept fixed and the attenuator rheostat is swept. Finally, this procedure is repeated for the bottom branch, with the top branch disconnected. In sharp contrast to $N^4$ or $2N^2$ measurements that would otherwise be needed, this approach requires only $4N$ measurements to cover the entire constellation. For the circuit described in Section~\ref{subsection_AC_circuit} that uses 8-bit rheostats with $N=2^{8}$, the results of the characterization measurements are given in Fig.~\ref{fig_algo}(a). The experimental details of these measurements are provided in Section~\ref{subsection_experiments} below. Here, by digitally controlling the rheostat states, a total of $4N = 2^{10}$ measurements were rapidly conducted (as opposed to $N^4 = 2^{32}$ or $2N^2 = 2^{17}$ measurements). 
Next, the unmeasured part of the constellation is estimated using the following equation:

\begin{equation}\label{eq:gain_eq}
	G_{vm}(x_1,x_2,x_3,x_4) = G_{1} (x_1,x_2) + G_{2}(x_3,x_4) 
\end{equation}
where
\begin{align}\label{eq:gain_eq2}
	G_{1}(x_1,x_2) &= G_{\phi_1} (x_1) \cdot G_{a_1}(x_2) \\
	G_{2}(x_3,x_4) &= G_{\phi_2}(x_3) \cdot G_{a_2} (x_4)
\end{align}
Here, $x_1, x_2, x_3, x_4 \in {1...N}$ are the wiper states determining the resistance value of the 4 rheostats, $G_{vm}$ is the estimated transfer function of the entire VM, and $G_{1}$ and $G_{2}$ are the estimated transfer functions of the top and bottom branches, respectively. In addition, $G_{\phi_1}, G_{a_1}$ and $G_{\phi_2}, G_{a_2}$ are the measured transfer functions of the phase shifter and attenuator in the top and bottom branches, respectively. A visual representation of the estimated constellation can be seen in Fig. \ref{fig_algo}(b). Finally, the top branch and bottom branch parts of the estimated constellation are stored in a look-up table (LUT). Note that this procedure only needs to be performed once for a given VM circuit.

\textit{2) Look-up Table Based AC Algorithm}: To compensate for the feedthrough that occurs during an actual experiment, we propose a LUT-based AC algorithm that can be divided into two parts: coarse adjustment and fine adjustment. In the coarse adjustment process, first, the initial direct feedthrough is measured with the VM disabled. Then, the VM is initialized with fixed digital inputs and its output is measured. Using these two measurements, a complex-valued scaling factor that describes the amplitude and phase differences between the constellation in the LUT and the VM output is estimated, and the constellation is scaled accordingly. After that, the branch suitable for the coarse adjustment is chosen as the one with the smaller distance between the initial feedthrough and the mean phase of the constellation for that branch. In the chosen branch's constellation, a compensation phasor is sought that will minimize the distance between the residual and the densely covered part of the other branch's constellation, under the constraint that the residual will lie in the support of the other branch's constellation. The rheostat wipers for the coarse adjustment branch are updated accordingly.

Next, a fine adjustment process is performed. First, the residual feedthrough after the coarse adjustment (i.e., the coarse residual) is measured. Note that it could also be estimated from the previous step using the LUT and the known initial feedthrough. Another search is performed in the constellation in which the coarse residual lies, to seek a compensation phasor that minimizes the final residual. Then, the rheostat wipers for the fine adjustment branch are updated accordingly.
The graphical demonstrations of the coarse and fine adjustments can be seen in Fig. \ref{fig_algo}(b) and (c), respectively. 

After applying the LUT algorithm, if the resulting residual feedthrough is not satisfactory, an iterative local search can be performed in the vicinity of the resulting residual. In this work, this local search is done by incrementing or decrementing each rheostat wiper one-by-one and measuring the new residual until either another minimum for the residual is reached or the residual starts to increase. This version of the algorithm is referred to as LUTit, where the appended "it" indicates the additional iterative local search. 

\subsection{Simulations}
Simulation studies were performed in phasor domain to demonstrate the performance of the VM circuit and the proposed AC algorithm. First, the frequency response of the VM circuit was simulated in SPICE. The AC algorithm simulations, performed in MATLAB, utilized the experimentally measured and estimated LUT (see Fig.~\ref{fig_algo}), which was assumed to be ideal and noise-free. First, Monte Carlo simulations were performed with 5000 repetitions. For each repetition, LUT algorithm ran with random feedthrough inputs, with independent and uniformly distributed real and imaginary parts. The cancellation performance was first evaluated without any noise, and then with white Gaussian noise added to both the measurements and the LUT at an interference-to-noise ratio (INR) of 40~dB to mimic the INR observed in the experiments. Next, the cancellation performance was evaluated as a function of INR at 20 different INR levels ranging between 14-94~dB, with white Gaussian noise of matching standard deviations added to both the measurements and the LUT at each INR level. At each INR level, Monte Carlo simulations were performed with 1000 repetitions. 


\subsection{Experiments}
\label{subsection_experiments}

The proposed technique was demonstrated on an in-house arbitrary waveform magnetic particle spectrometer (MPS) device \cite{tay_high-throughput_2016}. Shown in Fig.~\ref{fig_diag}b, this MPS device featured a DF coil with a small inductance of 3.8~\uH with 0.8 mT/A sensitivity \cite{top_arbitrary_2019}. The gradiometer receiver coil had 8.8~\uH inductance and had a manual adjustment knob for fine-tuned decoupling of the direct feedthrough.

The signal at the LNA output (i.e., the difference between the received signal and the VM output) was digitized at 2~MS/s by a data acquisition card (National Instruments PCIe-6374), which also generated and sent the DF pulses to a power amplifier (AETechron 7224). Thanks to the low inductance of the DF coil, the power amplifier was directly connected to DF coil without the need for impedance matching. The DF was at $f_0$ = 10~kHz with amplitudes ranging between 5-22.5 mT, depending on the experiment performed. Finally, the VM (see Fig.~\ref{fig_diag}c) was powered by batteries to ensure that no external interference coupled from the powerline to the received signal through the circuit. The entire setup (both the MPS and the VM) were controlled via MATLAB. 

With this experimental setup, 4 different types of experiments were performed: system characterization experiments, AC algorithm performance experiments, dynamic AC experiments, and MNP detection sensitivity experiments. All experiments, except for the sensitivity experiments, were performed with an empty MPS chamber.

\textit{1) System Characterization Experiments}: To create the LUT for AC, characterization experiments were performed using a DF with $f_0$ = 10~kHz frequency and 13.7~mT amplitude. This DF amplitude was chosen to maximize the VM input without heating the drive coil, to avoid rapid system drift. As described in Section~\ref{subsection_AC_circuit}, this DF was sensed by the pick-up coil and fed to the VM circuit. The VM output was measured using the single-ended mode of the LNA with an acquisition duration of 100~ms. Then, the attenuator and phase shifter rheostat wipers were swept one-by-one on the selected branch, while the other branch was disconnected using the disable function of the rheostats. Next, the output amplitude and phase of each measurement were estimated from the Fourier transforms of the acquired signals. This procedure was repeated for each branch and each rheostat one-by-one, resulting in $4N = 2^{10}$ measurements for the case of four 8-bit rheostats. Finally, the unmeasured phasors were estimated from these acquisitions using Eq. \ref{eq:gain_eq} to create a LUT with $2N^2 = 2^{17}$ entries. The entire characterization process took less than 5 minutes. 

\textit{2) AC Algorithm Performance Experiments}: The performances of the proposed LUT-based algorithms were compared with the following calibration-free nonlinear integer optimization algorithms: 
\begin{itemize}
\item Genetic Algorithm (GAL): This algorithm tries to minimize the direct feedthrough by applying a natural selection process \cite{salim_active_2018}. Briefly explained, GAL randomly generates candidate inputs for the VM, evaluates their performance at each iteration, and selects a subset of high performance inputs from which a new generation of candidate inputs are created. This process continues until convergence is achieved. For the experiments in this work, the population size for GAL was set as 10, which means that the algorithm generated 10 different candidates at each iteration, performed measurements of the residual to determine the cancellation performance of each candidate, and then selected the best ones to create the new generation of 10 candidates. 
\item Surrogate Optimization (SOP): This algorithm approximates the complex transfer function of the circuit in terms of simpler surrogate functions using several input/output measurements \cite{gutmann_radial_2001}. For the experiments in this work, SOP performed 8 measurements of the residual at the two-end wiper states of each rheostat to initialize the algorithm, followed by an iteration of measurements using the estimated optimal inputs. The approximated transfer function is updated at each iteration to determine the optimal inputs.
\end{itemize}

Experiments were performed to compare the cancellation performances and the execution times for the proposed LUT and LUTit algorithms, and the comparison methods GAL and SOP. Each algorithm was implemented in MATLAB to digitally and automatically control the rheostat states, without any manual intervention. During the experiments, 20 independent runs were conducted for each algorithm, with the VM states randomly reset at the beginning of each run. For all of the algorithms, 50~\ms~acquisition time was used for direct feedthrough measurements. For these experiments, the DF was at $f_0$ = 10~kHz with 6.8~mT amplitude. 

\textit{3) Dynamic AC Experiments}: The ability of the proposed method to dynamically reduce the direct feedthrough was demonstrated by inducing rapid system drift via applying high power to heat up the DF coil. Accordingly, a relatively high 22.5~mT amplitude DF was applied at $f_0$ = 10~kHz, with 100~ms pulses and 50~ms intervals throughout 20~seconds. During the first 5~seconds, the AC was kept off. Then, for the following 15~seconds, the direct feedthrough level was monitored during the 100~ms DF heating pulses. Whenever the feedthrough exceeded a threshold of -70~\si{\decibel\volt}, the algorithm kicked off automatically to reduce it down. The acquisition duration for monitoring and heating was set as 100~\ms, whereas the duration of each acquisition needed for the LUT or LUTit AC algorithms (see Section~\ref{subsection_AC_algo}) was set as 10~\ms.

To demonstrate the performance of the proposed AC method under different initial feedthrough levels, the above-mentioned experiment was repeated with 3 different initial cancellation levels by adjusting the knob of the gradiometer receive coil to different positions.

\textit{4) MNP Detection Sensitivity Experiments}:
To demonstrate the improvements achieved by the proposed AC method in an MPS setup, MNP samples were prepared using 3 different commercially available nanoparticles: Vivotrax (Magnetic Insight, USA), Resovist (Bayer, Germany), and Perimag (Micromod GmbH, Germany) with undiluted iron concentrations of 5.5~\mgpmL, 28~\mgpmL, and 25~\mgpmL, respectively. For each MNP, a dilution series at 9 different concentrations were prepared at [2, 10, 20, 50, 100, 200, 400, 1000] fold dilutions. With a total volume of 20~\si{\micro\liter} each, the resulting samples had total iron masses ranging between 0.05-250 $\mu$g. 
The DF was applied at $f_0 = 10~\kHz$ at 2 different amplitudes of 5~mT and 10~mT, where the system drift was relatively slow. For each vial, first, the LUTit version of the proposed AC algorithm was applied using 30~ms acquisitions. Then, the rheostat states were fixed to the optimal values determined by the algorithm. Next, 10 repeated empty-chamber baseline measurements were acquired, followed by 10 repeated measurements of the MNP signal. The acquisition times for the baseline signal and MNP signal were 20~ms each. This procedure was then repeated with the AC turned off. 

The signal strength at the harmonics, as well the root-mean-squared (RMS) value of the acquired signal were investigated with and without AC. Since the MNP signal at all harmonics should be proportional with the MNP concentration, $y=ax$ line was fit to the signal as a function of concentration for each nanoparticle type at each harmonic. The corresponding $R^2$ values were calculated to observe whether the measurements followed the expected linear trend. The baseline measurements were used to calculate the baseline signal level at each harmonic, which had contributions from both the direct feedthrough and noise. The signal levels that were at least one standard deviation (std) above the mean of the baseline were considered as being successfully detected.

\section{Results}

\begin{figure}[!t]
	\centering
	\includegraphics[width=\columnwidth]{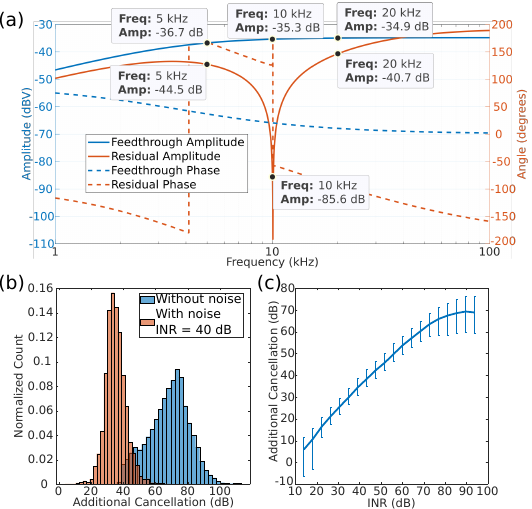}
	\caption{(a) Frequency response plots of cancellation and feedthrough, simulated by SPICE. Important values are annotated on the plots. (b) Histogram of the Monte-Carlo simulations with and without measurement noise added. (c) Median, 25th and 75th percentile of additional cancellation with respect to interference-to-noise (INR) ratio.}
	\label{fig_sim}
\end{figure}

The results of the VM circuit frequency response simulations are given in Fig.~\ref{fig_sim}(a). Comparing the feedthrough amplitude and the residual amplitude at the targeted $f_0 = 10~\kHz$, the proposed VM circuit can successfully achieve at least 40~dB cancellation (exceeding 50~dB at exactly $10~\kHz$) with 8-bit digital rheostats used in the design. Another observation from Fig.~\ref{fig_sim}(a) is, the bandwidth of the cancellation at $f_0 = 10~\kHz$ was quite narrow, with a 6~dB bandwidth of 1.7~kHz. This narrow bandwidth was expected due to the disparity between the phase responses of the VM and the direct feedthrough.

In Fig~\ref{fig_sim}(b), the results of the Monte-Carlo simulation with 5000 repetitions are plotted, showing the histogram of the additional cancellation provided by the LUT-based AC algorithm. In the absence of noise (i.e., at infinitely large INR), the algorithm provides 68$\pm$13~dB (mean$\pm$std) additional cancellation at $f_0$. Even in the presence of a realistic level of noise at INR = 40~dB, the algorithm provides 35$\pm$6~dB an additional cancellation at $f_0$. Next, Fig.~\ref{fig_sim}(c) shows the additional cancellation with respect to INR, where the error bars indicate the median and the $25^{th}-75^{th}$ percentiles of the Monte Carlo simulation results with 1000 repetitions at each INR level. As expected, the additional cancellation performance increases with INR, reaching a mean value of 67~dB at INR levels above 80~dB, where the interference strongly dominates over the noise. On the other hand, the performance deteriorates at low INR, slowly converging to an additional cancellation of 0~dB as INR approaches 0. Note that at INR = 0, there is no interference but only noise, which cannot be cancelled with or without AC. The additional cancellation level stays above 0~dB for INR$>$25~dB (corresponding to an interference that is at least 18 times larger than the noise), indicating that the proposed VM circuit and the LUT-based AC algorithm can ensure successful cancellation of the direct feedthrough in realistic MPI settings.

\begin{figure}[!t]
	\centering
	\includegraphics[width=\columnwidth]{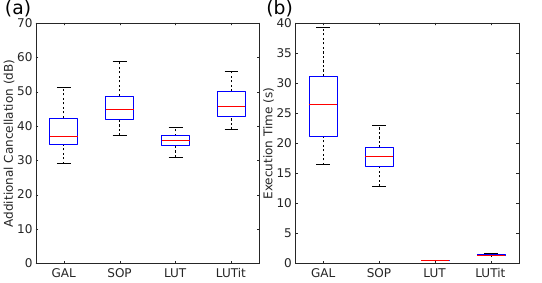}
	\caption{Comparison of different algorithms in terms of (a) additional cancellation provided and (b) execution time per run. The red lines show the median values, the boxes show 25$^{th}$-75$^{th}$ percentile values, and the whiskers show the extents of the measurements. }
	\label{fig_comp}
\end{figure}

The results for the VM system characterization experiments were already presented in Fig.~\ref{fig_algo}. Accordingly, Fig.~\ref{fig_algo}a shows that rheostats 1 and 3 (i.e., the phase shifters in the top and bottom branches, respectively) both have flat amplitude responses and provide 0 to $-\pi$ and 0 to $\pi$ phase coverages, respectively. On the other hand, rheostats 2 and 4 (i.e., the attenuators in the top and bottom branches, respectively) both have flat phase responses with over 50-fold variation in their amplitude responses. From these $4N = 2^{10}$ measurements, the estimated LUT constellation with $2N^2 = 2^{17}$ entries is demonstrated in Fig.~\ref{fig_algo}b, showing that the VM ensures sufficient and dense coverage of a range of phases. 

The experiment results comparing the performances of the proposed LUT and LUTit algorithms with SOP and GAL are shown in Fig.~\ref{fig_comp}, for 20 independent runs of each algorithm. As seen in Fig.~\ref{fig_comp}(a), LUTit had the highest additional cancellation performance among the tested algorithms. GAL was slightly better than LUT, whereas SOP outperformed both LUT and GAL. The additional cancellation levels were 39$\pm$6~dB for GAL, 46$\pm$6~dB for SOP, 36$\pm$3~dB for LUT, and 47$\pm$5~dB for LUTit (mean$\pm$std). By comparing the interference level at $f_0$ with the noise at non-harmonic frequencies, the INR level of these experiments was estimated as approximately 45~dB. Accordingly, we can conclude that the additional cancellation provided by LUTit in these experiments agrees well with the simulation results in Fig~\ref{fig_sim}(c). 

Next, comparing the execution times of the AC algorithms in Fig.~\ref{fig_comp}(b), both of the proposed algorithms were significantly faster than SOP and GAL. LUTit and LUT were at least 8-fold and 26-fold for faster than SOP, respectively. The execution times were 28$\pm$8~s for GAL, 18$\pm$4~s for SOP, 0.5$\pm$0.03~s for LUT, and 1.4$\pm$0.2~s for LUTit (mean$\pm$std). This major improvement in execution time for the proposed AC algorithms is due to the system characterization based approach adopted in this work. 
Overall, the results in Fig.~\ref{fig_comp} clearly demonstrate the superior performance of the LUT-based algorithms, as they provide competitive cancellation levels at significantly reduced execution times. The results also indicate that, LUTit should be preferred over LUT in the cases where the system drift is slow and a longer execution time can be allocated, as it provides improved cancellation performance. 

\begin{figure}[!t]
	\centering
	\includegraphics[width=\columnwidth]{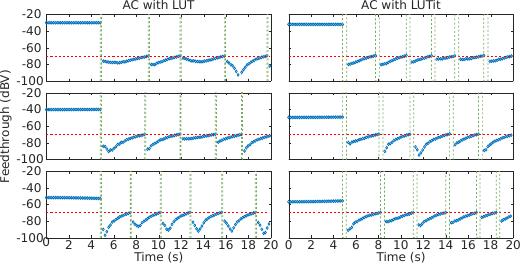}
	\caption{Results of the heating experiments. Green dashed boxes marks the time intervals where the AC is running. Red dashed line shows the threshold -70~dBV. AC is activated at the 5~second mark.}
	\label{fig_heat}
\end{figure}

Figure~\ref{fig_heat} shows the results of three different dynamic AC experiments for LUT and LUTit. Here, the gradiometer knob of the MPS receive coil was adjusted to a different position at the beginning of each experiment, as reflected with the differences in the initial feedthrough levels across experiments. Here, due to the heating of the DF coil under a relatively large 22.5~mT DF amplitude, the direct feedthrough drifts over time relatively quickly. Despite the rapid system drift, both of the proposed AC algorithms successfully keep the feedthrough under the desired threshold of -70~\si{\decibel\volt}. The algorithm runtimes for each cancellation cycle were measured as 100~ms for LUT and 340~ms for LUTit, respectively, with the additional time stemming from the local search iterations for LUTit. Note that LUTit did not provide an apparent increase in the additional cancellation, because the rapid drift of the direct feedthrough throughout the increased runtime of LUTit caused the chosen solution to diverge from the optimal solution. In fact, if the drift occurs extremely rapidly, LUTit can potentially perform worse than LUT for the same reason. In contrast, the drift was relatively slow for the results in Fig.~\ref{fig_comp}, where the system temperature remained constant and LUTit provided 11~dB additional cancellation over LUT on average.

\begin{figure*}[!t]
	\centering
	\includegraphics[width=0.9\textwidth]{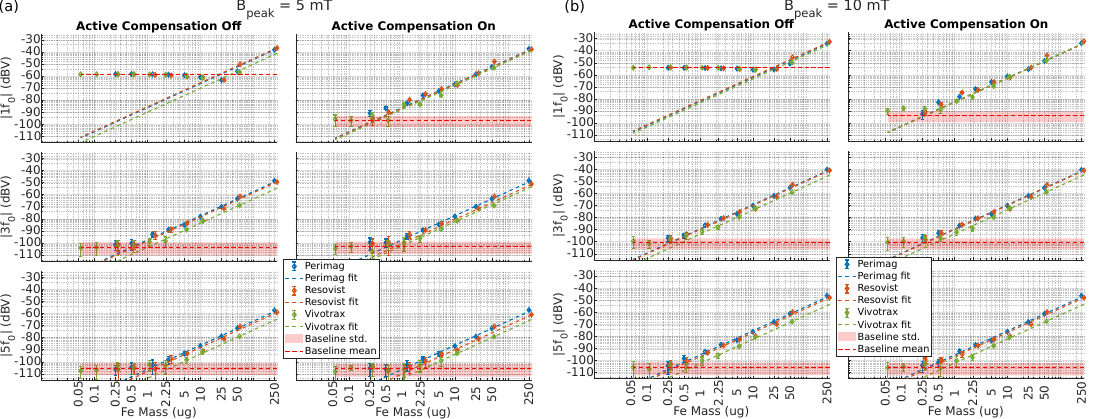}
	\caption{Magnitude of the measured signal's fundamental frequency, 3\textsuperscript{rd} and 5\textsuperscript{th} harmonics for each dilution and particle when (a) 5~mT and (b) 10~mT DF is applied.}
	\label{fig_mps}
\end{figure*}

\begin{figure*}[!t]
	\centering
	\includegraphics[width=0.9\textwidth]{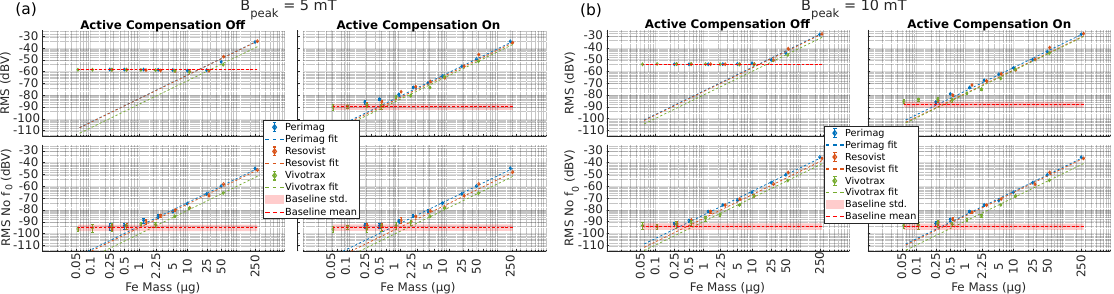}
	\caption{RMS magnitude of the measured signal and RMS when $f_0$ is filtered for (a) 5~mT and (b) 10~mT DF is applied.}
	\label{fig_rms}
\end{figure*}

Finally, the results of the MNP detection sensitivity experiments are given in Figs.~\ref{fig_mps}-\ref{fig_rms}. In Fig.~\ref{fig_mps}, the acquired signal strength at $f_0$, $3f_0$ and $5f_0$ are plotted as a function of the iron mass in the MNP sample at 5~mT and 10~mT DF amplitudes for 3 different nanoparticle types, without and with AC. The mean and the std of the baseline measurements are also marked. Likewise, Fig.~\ref{fig_rms} shows the RMS value of the acquired signal up to the 5th harmonic with and without including the signal contribution at $f_0$. $R^2$ values of the line fits were higher than 0.99 for the fundamental frequencies where AC was employed, verifying that the linear trend is preserved. Finally, in Table~\ref{table_xcross}, Fe masses where the line fits cross with baseline mean is shown to determine the minimum detectable Fe mass for each case. By looking at the RMS signal when AC is on, compared to AC off filtered RMS (RMS-$f_0$) case, up to 2.8 and 1.6-fold increase is observed for 5~mT and 10~mT~DF, respectively.

\begin{table*}[t]
	\caption{Detectable Fe masses (\ug).}
	\label{table_xcross}
	\centering
	\begin{tabular}{lc|ccl|ccl|ccl}
		&			& \multicolumn{3}{c}{Vivotrax}          & \multicolumn{3}{c}{Resovist} 	 	        & \multicolumn{3}{c}{Perimag}        \\
		&			& AC Off 	& AC On  &	Improvement	    & AC Off 	& AC On     &	 Improvement	& AC Off 	& AC On & Improvement    \\
		\hline
		\multirow{5}{*}{\begin{sideways}5 \mT\end{sideways}} 
		& RMS 		& 30    	&  0.67  & $\times$  44.8	& 17.49		&  0.56 	& $\times$ 31.2		& 17.19 	& 0.49  & $\times$ 35.1  \\
		& RMS-$f_0$	&  1.92 	&  2.02  & $\times$   0.95	&  1.04 	&  1.35		& $\times$  0.83	&  0.90  	& 0.92  & $\times$  0.98 \\
		&$|f_0|$ 	& 38.5 		&  0.36	 & $\times$ 107		& 23.02 	&  0.32 	& $\times$ 71.9		& 24.8 		& 0.31  & $\times$ 80    \\
		&$|3f_0|$ 	&  1.04  	&  0.99  & $\times$   1.05 	&  0.57 	&  0.79 	& $\times$  0.72	&  0.51 	& 0.56  & $\times$  0.91 \\
		&$|5f_0|$ 	&  2.72  	&  2.59	 & $\times$   1.05 	&  1.43 	&  1.65 	& $\times$  0.87	&  1.17 	& 1.08  & $\times$  1.08 \\
		\hline
		\multirow{5}{*}{\begin{sideways}10 \mT\end{sideways}} 
		& RMS 		& 20.35 	& 0.38   & $\times$  53.5	& 14.55 	& 0.36 		& $\times$  40.4	& 13.76 	& 0.28  & $\times$ 49.1  \\
		& RMS-$f_0$	&  0.64 	& 0.62 	 & $\times$   1.03	&  0.47 	& 0.37 		& $\times$ 	 1.27	&  0.34 	& 0.34  & $\times$  1    \\
		&$|f_0|$ 	& 29.66 	& 0.28 	 & $\times$ 105.9	& 23.64 	& 0.29 		& $\times$	81.5	& 25.51  	& 0.28  & $\times$ 91.1  \\
		&$|3f_0|$ 	&  0.46   	& 0.45 	 & $\times$   1.02	&  0.29 	& 0.29 		& $\times$	 1  	&  0.28 	& 0.28  & $\times$  1    \\
		&$|5f_0|$ 	&  0.72   	& 0.67 	 & $\times$   1.07	&  0.38 	& 0.36 		& $\times$	 1.05	&  0.32 	& 0.31  & $\times$  1.03 \\
	\end{tabular}
\end{table*}

\section{Discussion}

In this work, we proposed an active compensation method that can reduce direct feedthrough to the point where the fundamental frequency of highly diluted particles can be recovered. This allowed us to detect such particles, or increased the total signal acquired from these particles, which in turn increased the SNR.

Simulations in Fig. \ref{fig_sim} suggests that, the proposed vector modulator together with the proposed algorithm is able to reduce the direct feedthrough consistently and in a variety of INR conditions. By looking at the INR versus additional cancellation plot in Fig. \ref{fig_sim}(c), we can also say that the proposed method is suitable where the feedthrough is at least 25~dB above the noise floor.

Algorithm comparison study shows that, even though GAL and SOP were able to provide at least 30 and 40 dB cancellation, respectively, in regard to the execution times, they were significantly slower than the proposed algorithms LUT and LUTit. In cases where near real-time operation is desired, or scan time can be an issue (e.g. imaging), both LUT and LUTit can be used without sacrificing much additional isolation.

In the case that a calibration-free algorithm is needed, SOP is more suitable for the task rather than GAL, as SOP provides both higher additional cancellation and lower execution time than GAL. It should be noted that the parameters of SOP and GAL such as the population size and initial population range for GAL were set so that they would work well for different scenarios. More careful choice of these parameters can increase the performance of these algorithms.

By heating experiments in Fig. \ref{fig_heat}, we have shown that how direct feedthrough drifts over time, and how our proposed method can keep the feedthrough under the threshold. Moreover, in this case, we did not observe any improvement over LUT when we used LUTit. This is expected, since with heating, we expect the feedthrough to change quickly over time. Hence, in the case that feedthrough is expected to vary quickly, LUT can be preferred over LUTit.

The algorithm runtime was dependent on the acquisition time. On average, it was measured as 100~ms and 330~ms for LUT and LUTit, respectively when the acquisition time for AC was 10~ms and, 480~ms and 1.4~s for LUT and LUTit respectively when the acquisition time for AC was 50~ms. The execution times are bottle-necked by the four acquisitions that are used to measure the feedthrough, system gain, coarse and fine compensation results. The execution time can be reduced by reducing the overhead associated with these acquisitions.

In this paper, we only focused on the direct feedthrough cancellation at a single frequency, i.e., the fundamental harmonic. However, any source of nonlinearity in the transmit chain can also lead to feedthrough at the higher harmonics, especially if sufficient passive cancellation can not be achieved by the gradiometer receive coil. Although the proposed AC system can slightly decrease the feedthrough at the harmonics ($\sim 6 \text{ dB}$), it may not be sufficient if the feedthrough at the higher harmonics is substantial. For such cases, increasing the bandwidth can increase the practicality of the proposed device significantly. The main reason behind the narrow bandwidth is the disparity between the phase responses of the vector modulator and the feedthrough signal. One possible solution to this issue is the addition of a phase shifter after the receive coil in order to control the phase response of the received signal. However, this circuit would need to be ultra-low noise to not degrade the SNR of the received signal. Furthermore, the warped phase response of the particle signal due to the phase shifter would need to be corrected during postprocessing.

Although not investigated in this work, another issue that needs to be considered for the future work is the added noise coming from the vector modulator itself. For this experiment configuration, slight ($\sim 0.5 \text{ dB}$) increase of the noise floor was observed, which was deemed negligible. However, if a system with higher sensitivity is used (e.g., an LNA with lower noise and higher gain, or less noise coming from the receive coil), the vector modulator may become the dominant source of noise. Fortunately, noise from the vector modulator can be easily reduced by employing lower noise Op-Amps, or by designing the circuit in a noise-optimized fashion. Also, addition of an LNA before the vector modulator can readily solve the noise problem.

Our system provides robustness to time-varying changes on the direct feedthrough, especially if they are coming from the transmit side, as we sense the compensation signal from the transmit coil. However, our system is still sensitive to the changes occurring at the receive side, more prominently at the gradiometer. We have observed that, mechanical stability of the connecting cables and coils are essential to sustain the low feedthrough level. The necessity of removing the particle to measure the feedthrough level presents a challenge, as there is no straightforward way of observing whether the feedthrough increased meanwhile. Moreover, if the operator is not careful, inserting the sample into the coil is observed to increase the feedthrough significantly. Thus, to solve this issue, the coil and the connector cables need to be stabilized properly. Furthermore, a method to measure the feedthrough even if the particle is present in the coil could increase the practicality of the system significantly, especially for the longer scans.

In the future, phase shifters, or weighted summers can be replaced with their passive counterparts to reduce circuit complexity, and achieve better linearity. However, in this case, impedance of these passive filters needs to be taken care of. Usage of Op-Amp based filters allowed us to design the filter without considering the impedance matching and power losses.

To increase the additional cancellation, the number of bits in the rheostats can be increased, or more branches can be added. Adding more branches would increase the degree-of-freedom when covering the constellation, however, it would also increase the complexity. Furthermore, the proposed LUT-based algorithms would need to be modified, as they currently assume that the constellation is covered in two opposing parts.

A better way of increasing the cancellation performance would be optimizing the circuit for a specific coil. Especially, if the system will be applied to a fixed setup where the coils are not expected to move, such as an MPI scanner, we expect the feedthrough to wander in a small section of the constellation. Hence, the rheostat values and the capacitance values of the phase shifters can be chosen so that, the relevant section of the constellation is more densely sampled. This can also allow us to decrease the number of bits to achieve the same additional cancellation performance.   

The search that performed after LUT-based algorithm completes is currently a simple search that does not utilize the phase and amplitude information of the residual feedthrough. If a better search is implemented, that only iterates towards to minimize the residual feedthrough could dramatically decrease the required number of iterations, hence close the gap between LUT and LUTit.

The proposed algorithm can be further improved, especially to achieve a better cancellation performance for the lower INR cases. For example, currently, which branch's constellation will be used is determined by the angular distance of the feedthrough to the mean angle of the constellation. However, especially for the low INR cases, this method fails and decides in favor of the wrong constellation. A support vector machine can be used to decide in which constellation the feedthrough falls into. Furthermore, currently coarse adjustment is done by minimizing the distance of the residual to a point where we expect the opposite constellation to be dense. Instead, we can choose the point where we want the coarse residual to be by also considering a density map of the opposite constellation.

\section{Conclusion}

An active compensation method utilizing a digitally controlled vector modulator and a lookup-table-based cancellation algorithm is proposed to solve the time-varying direct feedthrough problem in MPI. This method can reduce the feedthrough by more than 40 dB in less than half a second, recover the fundamental frequency of the MPI signal, and enable a significant improvement in sensitivity.


\ifCLASSOPTIONcaptionsoff
  \newpage
\fi



\bibliographystyle{IEEEtran}
\bibliography{IEEEabrv,bib/MPIPaper.bib}

%








\end{document}